# Inversion symmetry-broken tetralayer graphene probed by second harmonic generation

Wenqiang Zhou[1,2,3#], Jiannan Hua[2,3#], Naitian Liu[2,3], Jing Ding[2,3], Hanxiao Xiang[2,3], Wei Zhu[2,3], Shuigang Xu[2,3*]

[1] *School of Physics, Zhejiang University, Hangzhou, 310027, China*
[2] *Key Laboratory for Quantum Materials of Zhejiang Province, Department of Physics, School of Science, Westlake University, 18 Shilongshan Road, Hangzhou 310024, Zhejiang Province, China*
[3] *Institute of Natural Sciences, Westlake Institute for Advanced Study, 18 Shilongshan Road, Hangzhou 310024, Zhejiang Province, China*

#These authors contributed equally to this work.
*Correspondence to: xushuigang@westlake.edu.cn

**Symmetry breaking governs most fascinating phenomena in crystals, such as ferroelectricity, nonlinear optics, piezoelectricity, ferromagnetism, and superconductivity. In two-dimensional materials, a wide variety of tuning knobs presents extraordinary opportunities for engineering symmetry breaking, leading to the emergence and manipulation of novel physical properties. Recently, tetralayer graphene with ABCB stacking order is predicted to possess atypical elemental ferroelectricity arising from the symmetry breaking induced by its specific stacking configuration. Experimentally unveiling the stacking-order dependent symmetry in tetralayer graphene is crucial to understand the intricate properties in the emergent graphene allotropes. Here, we observe pronounced nonlinear optical second harmonic generation (SHG) in ABCB-stacked tetralayer graphene, but absent in both ABAB- and ABCA-stacked allotropes. Our results provide direct evidence of symmetry breaking in ABCB-stacked tetralayer graphene. The remarkable contrast in the SHG spectra of tetralayer graphene allows straightforward identification of ABCB domains from the other two kinds of stacking order and facilitates the characterization of their crystalline orientation. The employed SHG technique serves as a convenient tool for exploring the intriguing physics and novel nonlinear optics in ABCB-stacked graphene, where spontaneous polarization and intrinsic gapped flat bands coexist. Our results establish ABCB-stacked graphene as a unique platform for studying the rare ferroelectricity in non-centrosymmetric elemental structures.**

The abundant tuning knobs in two-dimensional (2D) materials endow them versatile properties[1,2]. Notably, by twisting two adjacent graphene layers with a controllable angle, series of exotic phenomena have been discovered, such as correlated states[3], unconventional superconductivity[4,5], ferromagnetism[6], quantum anomalous hall effect[7] and topological valley currents[8]. Stacking order is another knob which can significantly influence electronic and optical properties of 2D materials. In 2D magnets, stacking order determines their magnetic orders[9-11]. In few-layer graphene, exotic correlated phenomena were found in rhombohedral- (ABC-) stacked few-layer graphene, such as Mott insulator[12], ferromagnetism[13,14], superconductivity[15,16] and electronic phase separation[17].

When stacking two layers of graphene on top of each other without twist, the most stable stacking order is known as Bernal stacking. In this configuration, half of carbon atoms in the upper layer are positioned directly over atoms in the lower graphene layer, while the remaining half align over the center of hexagons. Another conceivable configuration is AA stacking, where the upper and lower layers are exactly aligned. However, this arrangement is highly unstable and is not observed in reality. In the context of multilayer graphene, various stacking configurations between the individual layers produce versatile allotropes. Owing to the distinct interlayer interactions and crystal symmetries, multilayer graphene featuring diverse stacking orders exhibits prominently contrasting electronic band structures. In the case of tetralayer graphene, there are three inequivalent stacking orders, each accompanied by its unique crystal symmetry, as depicted in Fig. 1a. The prevalent one is the Bernal (ABAB) stacking, which is a semimetal displaying gate-tunable band overlap between the valence and conduction bands[18,19]. Another stacking order in tetralayer is the rhombohedral (ABCA) stacking, which attracts intense interest due to the emergence of a surface flat band at low energy[20-22]. The third and rarest stacking, denoted as ABCB or equivalently ABAC stacking, constitutes a peculiar system that has only recently been observed experimentally[23]. As illustrated in Fig. 1a, both ABAB and ABCA stackings preserve inversion symmetry, which is broken in ABCB stacking. Notably, ABCB-stacked tetralayer graphene appears to be the thinnest form of graphene that exhibits both inversion and mirror symmetry breaking[24]. Based on its non-symmetric structure, recent theoretical predictions have postulated that ABCB-stacked tetralayer graphene may possess a spontaneous out-of-plane electric polarization, potentially even manifesting sliding ferroelectricity[25,26]. We conducted a thorough investigation of the band structure of ABCB-stacked tetralayer graphene, utilizing a tight-binding model. The calculated results, as presented in Fig. 1b, reveal that ABCB-stacked tetralayer graphene can be conceptually understood as a coupling result of a monolayer graphene and an ABC-stacked trilayer graphene. Therein, a locally flat band and a massive Dirac band coexist near K points at low-energy region. Large density of states predominantly concentrates at the first and third layers. The broken layer inversion symmetry further endows ABCB graphene with an intrinsic bandgap, calculated to be approximately 11 meV. In the vicinity of the band edges, notable van Hove singularities emerge, associated with high density of states[23,27]. The resultant presence of a flat valley band at low energy favors electron-electron interactions, providing a new platform for exploring correlated phenomena, such as magnetism and unconventional superconductivity.

Second-harmonic generation (SHG) spectroscopy, known for its exceptional sensitivity to crystal symmetry, has been demonstrated to be a powerful technique to probe non-centrosymmetric structures, relative orientation, and the nonlinear properties of 2D materials, including $MoS_2$[28-30], h-BN[31], trilayer graphene[32] and $CrI_3$[33]. This reliable and noninvasive technique proves particularly suited for the characterization of materials lacking inversion symmetry, thereby affording sensitivity to the stacking order in tetralayer graphene. Herein, we report the observation of pronounced SHG in ABCB-stacked tetralayer graphene, stemming from its inherent symmetry breaking. Whereas, neither ABAB- nor ABCA-stacked allotropes show observable SHG signals. The discernible contrast in SHG response with respect to stacking order stands as a potent tool to visualize ABCB stacking domains within tetralayer graphene, as well as to determine their crystalline orientation.

**Raman characterization**

In our experiment, we prepared multilayer graphene through the mechanical exfoliation of natural graphite flakes onto a SiO$_2$/Si substrate with an oxide thickness of 285 nm. Subsequent characterization was conducted employing optical techniques. The thickness of multilayer graphene was first identified by optical contrast and then checked by Raman spectra (see in Fig. S1). Figure 1c shows an optical image of an exfoliated tetralayer graphene. Notably, there is no observable optical contrast among the various stacking domains in the flake with the same layer number.

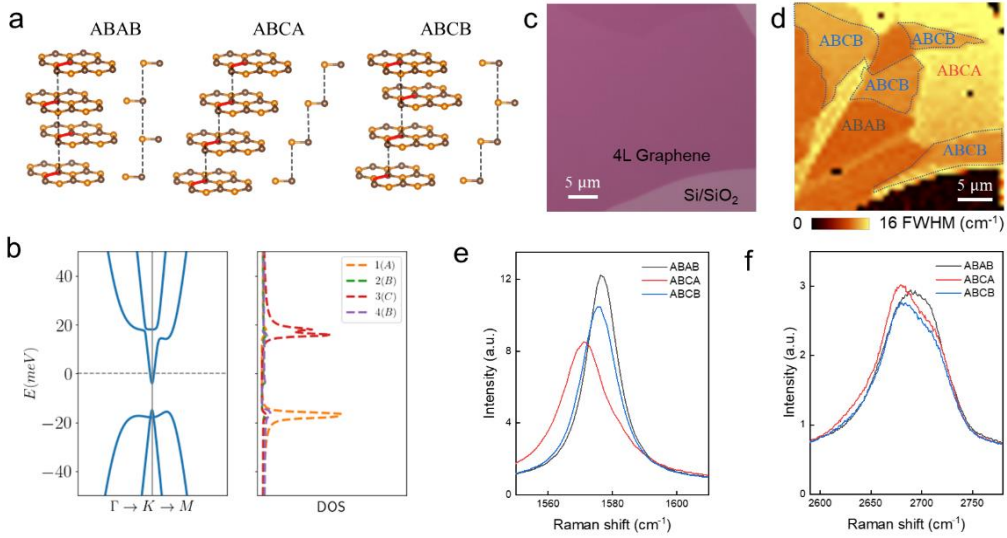

**Fig. 1| Band structure and Raman characterizations of tetralayer graphene allotropes. a**, Schematics of tetralayer graphene with ABAB, ABCA and ABCB stacking orders. The two equivalent carbon atoms representing A and B sublattices are distinctively colored. **b**, Calculated band structure (left) and density of states projected on each layer (right) of ABCB-stacked tetralayer graphene. **c**, Optical image of mechanically exfoliated few-layer graphene on a SiO$_2$/Si substrate with the oxide thickness of 285 nm. The tetralayer graphene with various stacking orders shares the identical optical contrast. **d**, Raman mapping of the FWHM of G band peak in the unit of cm$^{-1}$. The distinct contrasts distinguish three kinds of stacking orders. The corresponding optical image is in (**c**). **e,f**, Individual Raman spectra of the tetralayer graphene with three different stacking orders near G band peak from 1550 cm$^{-1}$ to 1610 cm$^{-1}$ (**e**) and 2D band peak from 2590 cm$^{-1}$ to 2780 cm$^{-1}$ (**f**). The excitation laser wavelength is 532 nm.

We first identified stacking domains within tetralayer graphene by Raman spectroscopy[34]. As shown in Fig. 1e and 1f, the Raman spectra of tetralayer graphene acquired at three different stacking domains manifest as distinct features in the peak positions of the G band and the line shapes of 2D band peak. Specifically, in ABCA stacking, a discernible redshift of approximately 3.2 cm$^{-1}$ relative to ABAB and ABCB stackings is evident in the G band peak position. This redshift originates from the slight difference of their phonon band structure. The 2D band peak in ABAB stacking exhibits a relatively symmetrical shape, whereas ABCA stacking displays a more pronounced asymmetric feature, characterized by a hump situated within the lower wavelength region, approximately around 2680 cm$^{-1}$. This observed difference in Raman spectra between ABAB- and ABCA- tetralayer graphene is similar to that observed in graphene with other layer numbers[34,35]. In the case of ABCB stacking, the G band peak has much closer characteristics to ABAB stacking, whereas the 2D band peak exhibits a feature between ABAB and ABCA stackings and makes itself reliable for domain identification. The origin of the 2D mode can be attributed to a double-resonance process, which renders it particularly sensitive to the electronic band structure of materials[36]. Consequently, the distinct 2D modes observed among various tetralayer graphene allotropes serve as indicators of stacking-dependent electronic band structure (elaborated further in Fig. S10). This finding suggests the potential presence of novel electronic properties within the recently unearthed ABCB graphene configuration, awaiting further exploration.

The multiple stacking domains within tetralayer graphene can be imaged through Raman maps employing metrics such as the full width at half maximum (FWHM) or the integrated intensity of the G band and 2D band. We have found Raman maps featuring the FWHM of the G band provide the highest contrast. Figure 1d illustrates a Raman map of tetralayer graphene (corresponding optical image shown in Fig. 1c), distinguishing three distinct stacking orders. It is worth noting that the relative FWHM values of the G band peak collected from ABAB and ABCB domains exhibit variability across different samples (see comprehensive discussion in Supplementary Section 2). Therefore, to accurately differentiate between ABAB and ABCB domains through Raman mapping utilizing the FWHM of the G band peak, a judicious approach should involve integrating it with the line shapes of the 2D band peak.

In this work, we have characterized a total of 35 tetralayer graphene flakes by Raman mapping, among which 5 flakes contain ABCB domains and 12 flakes contain ABCA domains. By collectively considering the overall area encompassed by the three distinct stacking domains across all flakes, we determined that the proportion of ABCA and ABCB domains are approximately 8% and 2%, respectively, consistent with previous report[23].

Notably, we observed that ABCB domains exhibit a higher level of instability when compared to ABCA domains, with a tendency to convert to alternative stacking orders. When illuminated with a high-power (>10 mW) laser beam for an adequate duration, both ABCB and ABCA domains shrink, and in some cases, completely convert to ABAB domains (see in Fig. S4). Occasionally, we even observed the phase transition from ABCA stacking to ABCB stacking (see in Fig. S4), a phenomenon suggestive of a surprisingly small energy barrier between these two stacking orders. We believe these transitions arise from the laser induced local heating effect. Similar domain motion under the influence of laser irradiation has also been observed in ABA/ABC stacked trilayer graphene[37]. The relatively low yield of ABCB stacking and its instability indicate its metastable character and explains why it escapes people's hunt over past decade until recently[23,34].

**Symmetry breaking in ABCB-stacked graphene**
While Raman spectra have proven valuable for detecting ABCB domains, they do exhibit limitations such as weakness, insensitivity, and high susceptibility to sample defects, disorders, and strains. In the subsequent discussion, we present compelling evidence that SHG serves as a more sensitive and effective method for identifying ABCB domains.

To this end, we conducted SHG spectroscopy on tetralayer graphene utilizing a reflection geometry with normal incidence excitation, as depicted schematically in Fig. 2a. As an initial demonstration of the remarkable utility of SHG spectroscopy in discerning the symmetry of graphene structures, we commenced by assessing the SHG in monolayer ($D_{6h}$ symmetry), bilayer ($D_{3d}$ symmetry), and trilayer ($D_{3h}$ symmetry) graphene.

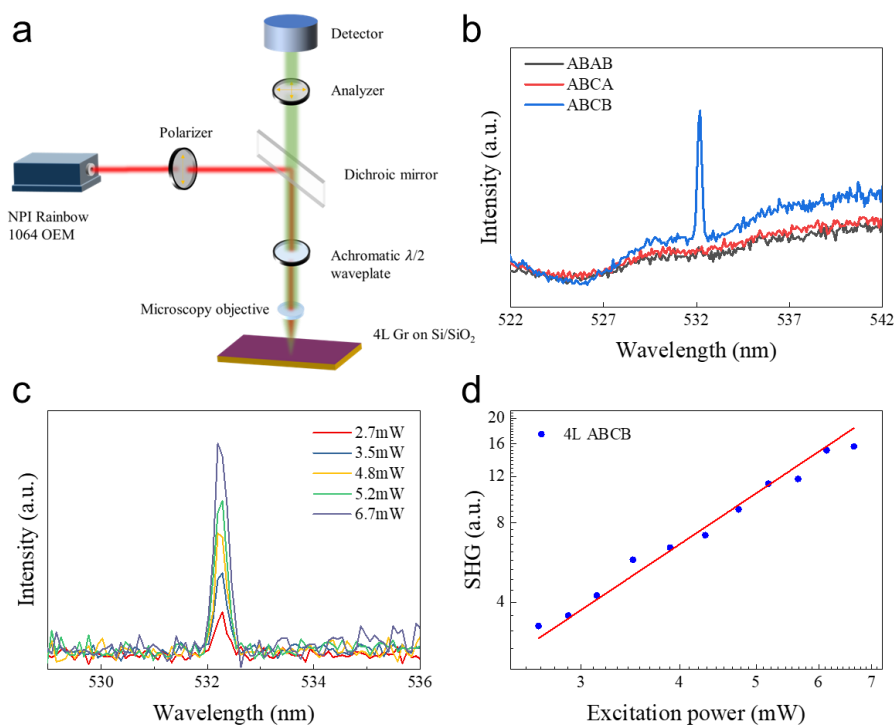

**Fig. 2| Second harmonic generation (SHG) spectroscopy of tetralayer graphene. a**, Schematic of optical setup utilized for the SHG measurement. **b**, The nonlinear spectra of tetralayer graphene with three kinds of stacking orders. Pronounced peak near 532 nm corresponding to the SHG signal appears in ABCB-stacked tetralayer graphene, whereas it is absent in ABAB and ABCA stacking. The stacking orders are initially determined based on Raman spectra. The excitation laser wavelength is 1064 nm. **c**, The excitation power dependence of SHG signal. **d**, The SHG peak intensity in (**c**) as a function of excitation power. The data are plotted in logarithmic coordinates, after subtracting the background. The solid line represents the anticipated quadratic dependence. Experimental data approximately conform to this quadratic trend, indicating the signal is indeed from second-order linearity described by electric dipole approximation.

Monolayer graphene is centrosymmetric with the inversion center at the center of the honeycomb lattice. Meanwhile, bilayer graphene with AB stacking configuration maintains its centrosymmetry. As a result, we observed an absence of SHG signal in monolayer and bilayer graphene as expected (see in Fig. S6). The symmetry of trilayer graphene depends on its stacking orders. We observed strong SHG signal in ABA-stacked trilayer graphene, a non-centrosymmetric configuration. Conversely, no SHG signal emerged from ABC-stacked trilayer graphene, distinguished by its centrosymmetric arrangement. These observations corroborate findings from previous studies, confirming the sensitivity of SHG spectroscopy in identifying the symmetry characteristics of graphene structures[32].

To detect the nonlinear features of tetralayer graphene, we subjected the sample to excitation by a 1064 nm pulse laser. In the investigation of a material's nonlinear properties, it is customary to utilize high excitation powers in order to elicit inherently weak nonlinear signals. To circumvent the possibility of laser-induced phase transitions in tetralayer graphene allotropes, as previously discussed, we meticulously conducted the SHG measurements within a controlled environment, maintaining a temperature of -30 °C under the protection of a dry nitrogen atmosphere. This condition also minimize potential disturbances from extraneous factors, such as charge doping (see the discussion in the Supplementary Section 4)[38].

Within the ABCB domain, we observed a prominent and narrow peak at 532 nm as shown in Fig. 2b, precisely half of the excitation wavelength, which is the hallmark of SHG signal. Importantly, this signal is conspicuously absent within both ABAB and ABCA domains. It is worth noting that for Bernal stacking graphene, evolving from trilayer to tetralayer, centrosymmetry is restored, coinciding with the disappearance of the SHG signal. To provide further validation of the SHG signal's authenticity, we explored the power dependence of the SHG signal within the ABCB domain. The results are shown in Fig. 2c. By effectively subtracting the broad background stemming from nonlinear photoluminescence signal[39], we proceeded to plot the relationship between SHG intensity and pump power. Figure 2d shows how the SHG intensity varies with excitation power, plotted in logarithmic coordinates. The solid line represents the anticipated quadratic dependence, characterized by a linear slope of 2. Notably, the experimental data closely approximate the quadratic relationship between SHG intensity and excitation power, which confirms the emission indeed arises from the SHG process—a phenomenon wherein two photons of the fundamental pump laser convert into one photon of the SHG signal. This agreement further indicates the simple electric dipole approximation can describe the SHG process in ABCB-stacked tetralayer graphene. This description can be expressed as $I_{\text{SHG}}(2\omega) \propto |\boldsymbol{P}(2\omega)|^2 \propto \left|\chi^{(2)}_{yyy} P(\omega)\right|^2$, where $I_{\text{SHG}}(2\omega)$ is the SHG intensity, $\boldsymbol{P}(2\omega)$ is SHG electric field vector, $P(\omega)$ is excitation power, $\chi^{(2)}_{yyy}$ is the second-order susceptibility, $\omega$ is the incident frequency[40].

Ascertaining the precise value of $\chi^{(2)}_{yyy}$ relies on the accurate measurements of SHG intensity, which is influenced by various experimental parameters including the optical alignment, detector efficiency, and laser frequency. To estimate $\chi^{(2)}_{yyy}$ of ABCB-stacked tetralayer graphene, we sought to establish a reference by comparing it with a well-studied and calibrated material—monolayer MoS$_2$. We measured the SHG signal of MoS$_2$ under the same condition. Based on the comparison of the SHG intensity data ($\chi^{(2)}_{yyy} = \chi^{(2)}_{\text{MoS}_2} \sqrt{I_{\text{SHG}}(2\omega)_{\text{Gr}}/I_{\text{SHG}}(2\omega)_{\text{MoS}_2}}$), we can deduce the sheet susceptibility of ABCB-stacked tetralayer graphene to be about $0.25 \times 10^4$ pm$^2$/V.

Given the significant disparity in SHG intensity among ABCB and the other two tetralayer graphene allotropes, we can utilize it to effectively map out ABCB domains. Figure 3c shows the mapping of integrated SHG intensity through sample scanning. Remarkably, the area characterized by pronounced SHG signal aligns with the ABCB domains identified in the Raman mappings. Nevertheless, compared with Raman mapping, SHG imaging obviously has much higher contrast due to the vanishing SHG in both ABAB and ABCA domains.

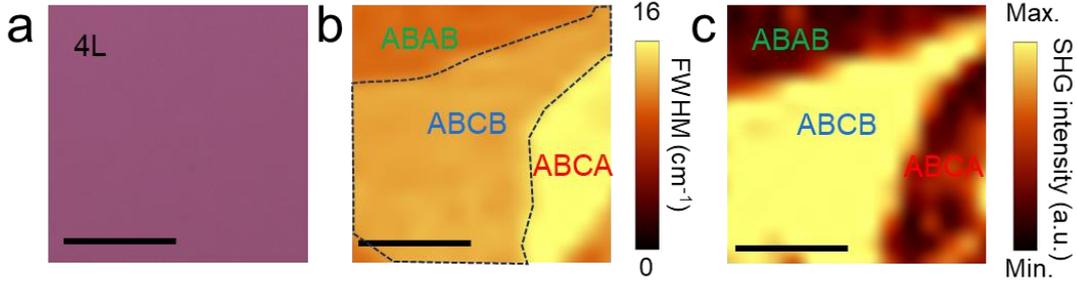

**Fig. 3| ABCB stacking domain probed by SHG imaging. a**, Optical image of a typical tetralayer graphene sample. **b**, Raman mapping of the FWHM of G band peak for the sample in (**a**). **c**, The corresponding SHG intensity mapping for the same region as that in (**b**). The scalar bars are all 5 μm.

**Polarization-resolved SHG in ABCB-stacked graphene**

SHG emerges as a versatile tool, not only enabling the identification of ABCB domains but also facilitating precise determination of the crystalline orientation. This capability can be realized by measuring polarization-resolved SHG. To this end, we insert a half-wave plate into the light path as shown in Fig. 2a to rotate the incident polarization vector with respect to the sample, which is controlled by a precise motorized rotator. The relative orientations of the incident and reflected lights are either in parallel or perpendicular, regulated by the polarizer and analyzer.

ABCB-stacked tetralayer graphene possesses a non-centrosymmetric $C_{3v}$ point group. Under normal incidence, when the polarizer and signal analyzer are aligned in parallel mode and perpendicular mode, the SHG intensity can be described as $I_{\parallel}(2\omega) \propto (\chi^{(2)}_{yyy})^2 \sin^2[3(\theta + \phi)]$ and $I_{\perp}(2\omega) \propto (\chi^{(2)}_{yyy})^2 \cos^2[3(\theta + \phi)]$, respectively (see the deduction process in the Supplementary Section 6)[41,42]. Here, $\chi^{(2)}_{yyy}$ is the $\chi^{(2)}$ susceptibility tensor element, $\phi$ is the angle between the polarization direction of the incident light and the zigzag direction of graphene, $\theta$ is the azimuthal angle controlled by rotating the half-wave plate in Fig. 2a. Figure 4a and 4b shows the polarization-revolved SHG patterns when the polarization of the incident light and the signal beam are in parallel (XX) and perpendicular (XY), respectively. The SHG intensity periodically varies with azimuthal angle $\theta$, appearing as six-petal patterns. These six-fold symmetrical patterns can be aptly fitted with the function $\sin^2[3(\theta + \phi)]$ and $\cos^2[3(\theta + \phi)]$ for parallel and perpendicular polarization components, respectively. Notably, these patterns exhibiting six-fold rotational symmetry under normal incidence conditions, are in accordance with the lattice symmetry inherent in the $C_{3v}$ point group for ABCB-stacked tetralayer graphene.

In order to elucidate the relationship between SHG peak in Fig. 4a and the underlying crystalline orientation, we initially select a tetralayer graphene flake with both an ABCB domain and a straight edge. The polarization-dependent SHG pattern reveals that the maximum intensity for the parallel (XX) mode corresponds exactly with the straight edge evident in the corresponding optical image as shown through the comparison between Fig. 4a and Fig. 4c. This straight edge is expected to align with either the armchair or zigzag directions. To determine the chirality of the graphene edge, we resorted to Raman spectra by the D band peak. It has been reported that in graphene, the intensity of D band peak near the armchair edge is stronger than that near the zigzag edge[43]. Figure 4d shows a Raman map of D band intensity near the two edges. The edge exhibiting a pronounced D band peak can be attributed to the armchair direction, whereas the one with negligible D band peak is close to the zigzag direction. With this identification, we assign back to polarization-revolved SHG patterns in Fig. 4a. We can conclude that when the polarizer and signal analyzer are aligned in parallel, the direction exhibiting the maximum SHG signal corresponds to the armchair direction. This conclusion is further confirmed by comparing polarization-revolved SHG patterns of ABCB-stacked tetralayer graphene with well-studied patterns of ABA-stacked trilayer graphene in the same flake (see in Fig. S9).

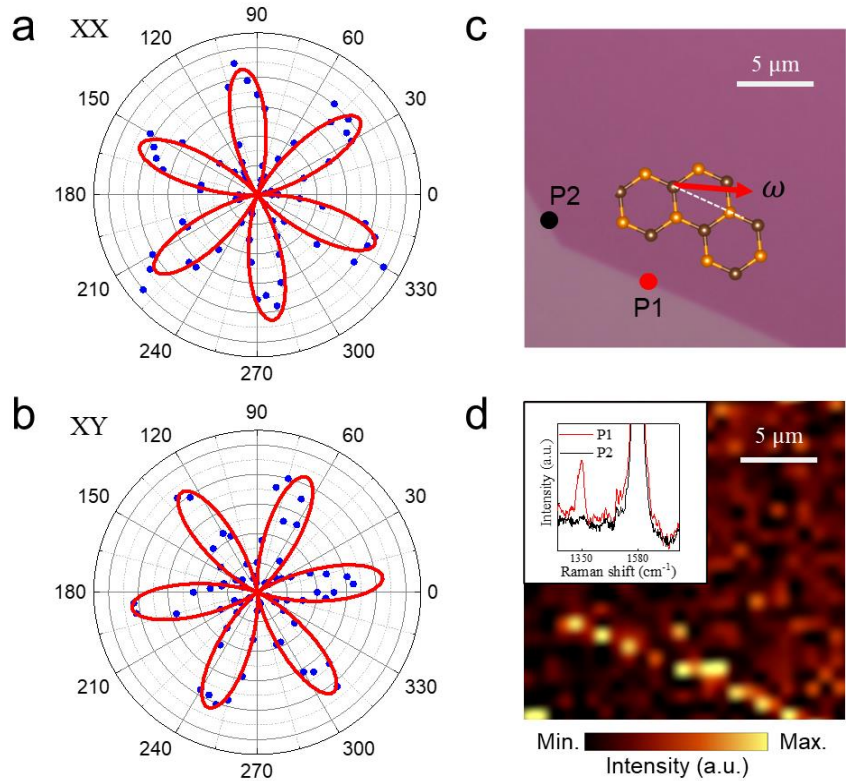

**Fig. 4| Crystalline orientation probed by polarization-resolved SHG. a,b**, Azimuthal angle-dependent patterns of SHG intensity within an ABCB-stacked tetralayer graphene, plotted in polar coordinates. The XX (**a**) and XY (**b**) panels show polarization-resolved SHG patterns, with the fast axis of the polarizer positioned in parallel and perpendicular to that of the analyzer, respectively. The red solid lines show the fitting results by the $I_0\sin^2[3(\theta+\phi)]$ and $I_0\cos^2[3(\theta+\phi)]$ curves for panel XX (**a**) and XY (**b**), respectively. **c**, The corresponding optical image of the sample. The SHG patterns in (**a**) and (**b**) indicate this straight edge aligns with the armchair direction of the graphene crystal structure, as illustrated in the schematics. **d**, Raman mapping of integrated D band peak intensity near the straight edges for the same sample in (**c**). Individual Raman spectra near D band are shown in the inset. The edge with a strong D band peak is believed to correspond to the armchair direction.

The polarization-dependent SHG measurements offer conclusive confirmation of the symmetry breaking in ABCB-stacked tetralayer graphene. Furthermore, the exhibited anisotropy within the SHG signal offers us with a precise means to determine the crystalline orientation of tetralayer graphene, which is quite useful in future's applications of this material, for instance, for fabricating ABCB tetralayer graphene/h-BN moiré superlattice.

**Outlook**

The unveiling of symmetry breaking in ABCB-stacked tetralayer graphene, as revealed by the SHG analysis in this work, stands as a pointer to the potential presence of spontaneous polarization or even ferroelectricity within this system[44]. Together with theoretically predicted emergence of narrow bands featuring an intrinsic gap due to the broken inversion and mirror symmetries, ABCB-stacked tetralayer graphene provides an outstanding platform for exploring the intricate interplay between correlated physics and ferroelectricity. Moreover, benefiting from the broad optical absorption and the electrically controllable optical response intrinsic to the graphene family, ABCB-stacked tetralayer graphene presents itself as a compelling optical material for designing nonlinear photonic and optoelectronic devices. Additionally, our work provides a highly sensitive tool for the identification of elusive ABCB domains within tetralayer graphene, distinguishing them from the other two companions, and concurrently facilitating the characterization of their crystalline orientation.

**Methods:**

**Flake preparation**

We prepared tetralayer graphene by mechanically exfoliating commercial graphite crystals (Graphenium Flakes, NGS Naturgraphit) on $SiO_2$/Si substrates with an oxide thickness of 285 nm. The layer numbers of graphene flakes were first identified by optical contrast using an optical microscope (Nikon LV100ND) and subsequently checked by Raman spectrum[45]. This process remained consistent with conventional procedures for graphene flake preparation, without any modifications. ABCB-stacked tetralayer graphene can be occasionally hunted by Raman spectroscopy. We also notice that ABCB stacking domains in tetralayer graphene usually coexists with ABCA stacking domains.

**Raman Spectroscopy**

Raman measurements were conducted using a commercial confocal Raman spectrometer (Witec Alpha 300RAS with UHTS 300 spectrometer) equipped with a 532 nm excitation laser. The spectra were initially calibrated by the 520 $cm^{-1}$ Raman peak of the silicon substrate. The laser beam's spot size was approximately 500 nm, achieved by using a 100x objective (NA=0.9). The scattered Raman signal was dispersed by a 1200 grooves per millimeter diffraction grating and detected by a silicon charge-coupled device. The excitation power was kept below 10 mW to avoid any domain motion or conversion due to the metastable characteristics of ABCB and ABCA stackings.

**Second harmonic generation**

SHG measurements were conducted utilizing the same experimental setup as that employed for Raman measurements. The sample was excited by a linearly polarized 1064 nm ultrafast fiber laser (Rainbow 1064, 15 ns, 10 MHz). The laser beam was focused by a 50x objective lens with a numerical aperture of 0.55 and shined onto the sample at normal incidence. The emitted signal was collected through the same lens in the reflection geometry and subsequently detected by the UHTS 300 spectrometer. The acquisition of SHG spectroscopy mappings was accomplished by systematically scanning the sample on a positioning stage. For the polarization-dependent SHG measurements, an achromatic half-wave plate was inserted between the dichroic beam splitter and the objective lens. This waveplate was driven by a motorized rotator. The emitted signal from the sample passed through the same waveplate and was analyzed by a linear polarizer before its entry into the spectrometer. All experiments were conducted within a nitrogen-purged sample holder to ensure a dry and controlled environment, to avoid the potential disturbing signal stemming from the external environment[38].

**Band structure calculation**

To consider the band structure and density of states of tetralayer graphene with ABAB, ABCA and ABCB stacking orders, we adopt the Slonczewski-Weiss-McClure (SWMC) tight-binding lattice model. The correction of the tight-binding Hamiltonian proposed by Jung et al.[46] is considered. The constructed SWMC tight-binding model allows us to perform the calculations of electric structure for multilayer graphene and the k-resolved contributions from each band. The band structure calculations are detailed in Supplementary Section 7.

**Acknowledgements:**

This work was funded by the National Natural Science Foundation of China (Grant No. 12274354), the Zhejiang Provincial Natural Science Foundation of China (Grant No. XHD23A2001), the R&D Program of Zhejiang province (2022SDXHDX0005) and the Westlake Education Foundation at Westlake University. The authors thank the Instrumentation and Service Centers for Molecular Science, the Instrumentation and Service Center for Physical Sciences (ISCPS) and the Westlake Center for Micro/Nano Fabrication at Westlake University for the facility support. The authors also thank Zhong Chen and Yuan Cheng for technical assistance in data acquisition and Wenjie Zhou for the fruitful discussion.

**Author contributions**

S.X. conceived the idea and supervised the project. W.Zhou prepared the samples with the assistance of N.L., J.D. and H.X.. W.Zhou performed the Raman and SHG measurements with help from N.L.. J.H. and W.Zhu performed the band structure calculations. S.X. and W.Zhu wrote the paper with the input from W.Zhou and J.H.. All authors contributed to the discussions.